\documentclass{article}
\usepackage{amsfonts}
\usepackage{amsmath}
\usepackage{amssymb}
\usepackage{graphicx}
\begin{document}

\title{ Rotating Frames in \emph{SRT}: Sagnac's Effect and Related Issues%
\thanks{%
published in \emph{Found. Phys.} \textbf{31}, 1767-1783 (2001).}}
\author{\quad \ W. A. Rodrigues, Jr.$^{(1,2)}\hspace{0.01in}\thanks{%
e-mail: walrod@ime.unicamp.br }\hspace{0.01in}$ and M. Sharif$^{(1)}$
\thanks{%
Permanent Address: Department of Mathematics, Punjab University,
Quaid-e-Azam Campus Lahore-54590, PAKISTAN, e-mail: hasharif@yahoo.com} \\
%EndAName
$^{(1)}${\footnotesize Institute of Mathematics, Statistics and Scientific
Computation }\\
{\footnotesize \ IMECC-UNICAMP CP 6065}\\
{\footnotesize \ 13083-970 Campinas, SP, Brazil}\\
{\footnotesize \ and}\\
{\footnotesize \ }$^{(2)}${\footnotesize Wernher von Braun}\\
{\footnotesize \ Advanced Research Center, UNISAL}\\
{\footnotesize \ Av. A. Garret, 257}\\
{\footnotesize \ 13087-290 Campinas, SP Brazil}\\
pacs: 04.90+e \ 03.30+p}
\date{05/10/2001}
\maketitle
\tableofcontents

\begin{abstract}
After recalling the rigorous mathematical representations in Relativity
Theory (\emph{RT}) of (i): observers, (ii): reference frames fields, (iii):
their classifications, (iv) naturally adapted coordinate systems (\emph{nacs}%
) to a given reference frame, (v): synchronization procedure and some other
key concepts, we analyze three problems concerning experiments on rotating
frames which even now (after almost a century from the birth of \emph{RT})
are sources of misunderstandings and misconceptions. The first problem,
which serves to illustrate the power of rigorous mathematical methods in
\emph{RT }is the explanation of the Sagnac effect (\emph{SE}). This
presentation is opportune because recently there are many non sequitur
claims in the literature stating that the \emph{SE} cannot be explained by
\emph{SRT}, even disproving this theory or that the explanation of the
effect requires a new theory of electrodynamics. The second example has to
do with the measurement of the one way velocity of light in rotating
reference frames, a problem for which many wrong statements appear in recent
literature. The third problem has to do with claims that only Lorentz like
type transformations can be used between the \emph{nacs }associated to a
reference frame mathematically moddeling of a rotating platform and the
\emph{nacs} associated with a inertial frame (the laboratory). Whe show that
these claims are equivocated.
\end{abstract}

\section{Introduction}

In this paper after recalling the rigorous mathematical definitions of
observers, reference frames, naturally adapted coordinate system (\emph{nacs}%
) to a given reference frame, the synchronization procedure and some other
related issues according to the formalism of \emph{RT} we show that: (i)
Contrary to recent claims [1-6], Sagnac's effect [7] has a trivial
explanation within \emph{RT}\footnote{%
This fact is clear from the the very well known paper by Post [8] and also
from the more recent references [10-14]. Even if all these references are
very good we think that there is space for a new presentation of the subject
using the modern concept of \emph{reference frames} \emph{fields}, because
such a presentation permit us to clarify some other obscures issues
concerning rotating reference systems (see sections 3.1 and 3.2).\emph{\ }\ }%
; (ii) We discuss also some problems connected with the standard
synchronization of standard clocks at rest on reference frames in
roto-translational motion and what can be expected for the measurement of
the one way velocity of light in such a frame. We show that our results are
at variance with statements found e.g., in [1,9]; (iii) We investigate also
claims [16-19] that the naturally adapted coordinate system (\emph{nacs}) to
the reference frame modelling a rotating platform must be related with the
\emph{nacs} associated with an inertial frame modelling the laboratory by a
Lorentz like type transformation [see eq.(ref{31}) later], showing that
these claims are without any physical or mathematical meaning.

\section{Some Basic Definitions}

Let $\tau$ be a spacetime theory as defined in [19, 24]. Let $Mod\tau$ be
the class of all models of $\tau$. Theory $\tau$ is said to be relativistic
theory if each $\tau\in ModT$ contains a substructure $ST=<M,D,\mathbf{g}>$
that is relativistic spacetime as defined in [19,24]. We remember here that $%
\mathbf{g}$ is a Lorentz metric and $D$ is the Levi-Civita connection of $%
\mathbf{g}$ on $M$.

\textbf{Definition 1.} Let $ST$ be a relativistic spacetime. A \emph{moving
frame} at $x\in M$ is a basis for the tangent space $T_{x}M$. An \emph{%
orthonormal frame} for $x\in M$ is a basis of orthonormal vectors for $%
T_{x}M $.

\textbf{Proposition 1.} Let $\mathbf{Q}\in\sec TU\subset\sec TM$ be a
time-like vector field such that $\mathbf{g}(\mathbf{Q},\mathbf{Q})=1$.
Then, there exist, in a coordinate neighborhood $U$, three space-like vector
fields which together with $\mathbf{Q}$ form an orthogonal moving frame for $%
x\in U$. The proof is trivial [25].

\textbf{Definition 2.} A non-spinning particle on $ST$ is a pair $(m,\sigma)$
where $\sigma:\mathcal{R}\supset I\rightarrow M$ is a future pointing causal
curve [19-22] and $m\in[0,+\infty)$ is the mass. When $m=0$ the particle is
called a photon. When $m\in(0,+\infty)$ the particle is said to be a
material particle. $\sigma$ is said to be the world line of the particle.

\textbf{Definition 3.} An \emph{observer} in $<M,D,\mathbf{g}>$ is a future
pointing time-like curve $\gamma:\mathcal{R}\supset I\rightarrow M$ such
that $\mathbf{g}(\gamma_{*}u,\gamma_{*}u)=1$. The inclusion parameter $%
I\rightarrow\mathcal{R}$ in this case is called the proper time along $\gamma
$, which is said to be the world line of the observer.

\textbf{Observation 1.} The physical meaning of proper time is discussed in
details, e.g., in [20,21] which deals with the theory of time in
relativistic theories.

\textbf{Definition 4.} An \emph{instantaneous observer} is an element of $TM$%
, i.e., a pair $(z,\mathcal{Z}),$ where $z\in M$, and $\mathcal{Z}\in T_{z}M$
is a future pointing unit time-like vector.

The Proposition 1 together with the above definitions suggests:

\textbf{Definition 5.} A \emph{reference frame} in $ST=<M,D,\mathbf{g}>$ is
a time-like vector field which is a section of $TU,U\subseteq M$ such that
each one of its integral lines is an observer.

\textbf{Observation 2.} In [19,22] an arbitrary reference frame $\mathbf{%
Q\in \sec}TU\subseteq\sec TM$ is classified as (i), (ii) below.

(i) according to its \emph{synchronizability}$.$ Let $\alpha_{\mathbf{Q}}=g(%
\mathbf{Q},)$. We say that \textbf{Q} is locally synchronizable iff $\alpha_{%
\mathbf{Q}}\wedge d\alpha_{\mathbf{Q}}=0$. \textbf{Q }is said to be locally
proper time synchronizable iff $d\alpha_{\mathbf{Q}}=0$. \textbf{Q }is said
to be synchronizable iff there are $C^{\infty}$ functions $h,t:M\rightarrow%
\mathcal{R}$ such that $\alpha_{\mathbf{Q}}=hdt$ and $h>0.$ These
definitions are intuitive.

(ii) according to the \emph{decomposition}\footnote{%
The validity of decomposition (1) is proved in Appendix C.} of
\begin{equation}
D\alpha_{\mathbf{Q}}=\mathbf{a}_{\mathbf{Q}}\otimes\alpha_{\mathbf{Q}}+%
\mathbf{\omega}_{\mathbf{Q}}+\mathbf{\sigma}_{\mathbf{Q}}+\frac{1}{3}\mathbf{%
\Theta}_{\mathbf{Q}}\mathbf{h},   \label{0}
\end{equation}
where
\begin{equation}
\mathbf{h}=\mathbf{g}-\alpha_{\mathbf{Q}}\otimes\alpha_{\mathbf{Q}}
\label{1}
\end{equation}
is called the projection tensor (and gives the metric of the rest space of
an instantaneous observer [20,22]), $\mathbf{a}_{\mathbf{Q}}$ is the (form)
acceleration of \textbf{Q}, $\mathbf{\omega}_{\mathbf{Q}}$ is the rotation
of \textbf{Q}, $\mathbf{\sigma}_{\mathbf{Q}}$ is the shear of \textbf{Q} and
$\mathbf{\Theta}_{\mathbf{Q}}$ is the expansion of \textbf{Q} . In a
coordinate chart ($U,x^{\mu}$), writing $\mathbf{Q}=Q^{\mu}\partial/\partial
x^{\mu}$ \ and $\mathbf{h}=(g_{\mu\nu}-Q_{\mu}Q_{\nu})dx^{\mu}\otimes
dx^{\nu }$ we have
\begin{align}
\mathbf{\omega}_{\mathbf{Q}\mu\nu} & =Q_{\left[ \mu;\nu\right] },  \notag \\
\mathbf{\sigma}_{\mathbf{Q}\alpha\beta} & =[Q_{\left( \mu;\nu\right) }-\frac{%
1}{3}\mathbf{\Theta}_{\mathbf{Q}}h_{\mu\nu}]h_{\alpha}^{\mu}h_{\beta }^{\nu},
\notag \\
\mathbf{\Theta}_{\mathbf{Q}} & =Q^{\mu};_{\mu}.   \label{2}
\end{align}

We shall need in what follows the following result that can be easily
proved:
\begin{equation}
\alpha_{\mathbf{Q}}\wedge d\alpha_{\mathbf{Q}}=0\Leftrightarrow\mathbf{%
\omega }_{\mathbf{Q}}=0.   \label{3}
\end{equation}

Eq.(3) means that rotating reference frames (i.e., frames for which $\mathbf{%
\omega}_{\mathbf{Q}}\neq0$) are not locally synchronizable and this result
is the key in order to solve the misconceptions usually associated with
rotating reference frames as the ones studied in section 3.

\textbf{Observation 3}. In Special Relativity where the space time manifold
is $<M\mathcal{=R}^{4},\mathbf{g}=\mathbf{\eta},D^{\mathbf{\eta}}>$\footnote{%
$\mathbf{\eta}$ is a constant metric, i.e., there exists a chart $\langle
x^{\mu}\rangle$ of $\ M=\mathcal{R}^{4}$ such that $\mathbf{\eta }%
(\partial/\partial x^{\mu},\partial/\partial x^{\nu})=\eta_{\mu\nu}$, the
numbers $\eta_{\mu\nu}$ forming a diagonal matrix with entries $(1,-1,-1,-1)$%
. Also, $D^{\mathbf{\eta}}$ is the Levi-Civita connection of $\mathbf{\eta}$.%
} an \emph{inertial reference frame }(\emph{IRF}) $\mathbf{I}\in\sec TM$ is
defined by $D^{\mathbf{\eta}}\mathbf{I}=0$. \

\textbf{Observation 4}. Before concluding this section it is very much
important to recall that a reference frame field as introduced above is a
\emph{mathematical} instrument. It did not necessarily need to have a
material substratum (i.e., to be realized as a material physical system) in
the points of the spacetime manifold where it is defined\footnote{%
More on this issue is discuted in [24].}. More properly, we state that the
integral lines of the vector field representing a given reference frame do
not need to correspond to worldlines of real particles. If this crucial
aspect is not taken into account we may incur in serious misunderstandings.
This observation will become clear in what follows.

\section{Rotating Reference Frames in \emph{SRT}}

In order to grasp the use of the very abstract notions introduced above we
shall study in this sections three problems. In section 3.1 we give an
analysis of the Sagnac effect [7], since there are many claims in the recent
literature saying that this effect cannot be explained by \emph{SRT} [1-3]
or even requires the formulation of a new electrodynamics where the photon
has a non zero mass [4]\footnote{%
Vigier [4] also claims that the Sagnac effect implies in the existence of a
fundamental frame.} or where the usual $U(1)$ gauge theory is substituted by
a $SU(2)$ [5] (or $O(3)$, [6]) gauge theory\footnote{%
The wrong claims of [1-3] have also been discussed and clarified in two
recent papers by Tartaglia and Rizzi [13] and by Matolcsi [14]. We decided
to present our analysis of the claims of [1-3] because we think that our
presentation has complementary material concerning the ones in [11,12] and
because it serves the proposal of showing how the concept of reference
frames as introduced here works. Moreover, we are not going to discuss the
claims of [4] and of [5,6] in this paper, leaving such an analysis for a new
paper. Here, we only call the reader's attention that the proposal of [6] is
almost identical to that of [5], which could eventually be a difficult
reference to be found.}. In section 3.2 we analyse the claims of [9] that
supposedly proved that clocks on a rotating platform that are synchronized
\`{a} l'Einstein at a given instant will not measure the standard value of
the \emph{one way} velocity of light after that the platform makes half a
rotation (as seen by an \emph{IRF}). In section 3.3 we analyze the claims of
[17-18] that it is necessary in order to describe a rotating platform,
coordinate functions necessarily different from the coordinate functions
defined by eq.(\ref{18}) later.

In this section $(M,\mathbf{g},D)$ denotes Minkowski spacetime and let $%
\mathbf{I\in}\sec TM$ be an inertial reference frame on $M$. Let $%
(t,r,\phi,z)$ be cylindrical coordinates for $M$ such that $\mathbf{I}%
=\partial/\partial t$ . Then
\begin{equation}
\mathbf{g}=dt\otimes dt-dr\otimes dr-r^{2}d\phi\otimes d\phi-dz\otimes dz
\label{12}
\end{equation}
Let $\mathbf{P}\in TM$ be another reference frame on $M$ given by
\begin{equation}
\mathbf{P}=(1-\omega^{2}r^{2})^{-1/2}\frac{\partial}{\partial t}+\omega
r^{2}(1-\omega^{2}r^{2})^{-1/2}\frac{\partial}{\partial\phi}   \label{13}
\end{equation}
\textbf{P} is well defined in an open set $U\subset M$ such that (Note that
the coordinate system $(t,r,\phi,z)$ is not submitted to the restriction $%
0<r<1/\omega$ of Eq.(\ref{16}))
\begin{equation}
U\equiv(-\infty<t<\infty;0<r<1/\omega;0\leq\phi<2\pi;-\infty<z<\infty)
\label{14}
\end{equation}
Then,
\begin{equation}
\alpha_{\mathbf{P}}=\mathbf{g}(\mathbf{P},)=(1-\omega^{2}r^{2})^{-1/2}dt-%
\omega(1-\omega^{2}r^{2})^{-1/2}d\phi   \label{15}
\end{equation}
and
\begin{equation}
\alpha_{\mathbf{P}}\wedge d\alpha_{\mathbf{P}}=-\frac{2\omega r}{(1-\omega
^{2}r^{2})^{2}}dt\wedge dr\wedge d\phi   \label{16}
\end{equation}
The rotation (or vortex) vector $[\Omega=\widetilde{\mathbf{g}}(*(\alpha _{%
\mathbf{P}}\wedge d\alpha_{\mathbf{P}}),)]$ is
\begin{equation}
\Omega=\omega(1-\omega^{2}r^{2})^{-1/2}\frac{\partial}{\partial z},
\label{17}
\end{equation}
where $\widetilde{\mathbf{g}}$ is the metric in the cotangent bundle. This
means that $\mathbf{P}$ is rotating with angular velocity $\omega$ (as
measured) according to $\mathbf{I}$ in the $z$ direction and local angular
velocity $\omega/(1-\omega^{2}r^{2})^{-1/2}$ as measured by an observer at
rest on $\mathbf{P}$ at a point with coordinates $(t,r,\phi,z=0)$. This
result is clear if we remember that a clock at rest on $\mathbf{P}$ at $%
(t,r,\phi ,z=0) $ has its period greater by a factor $(1-%
\omega^{2}r^{2})^{-1/2}$ in relation to clocks at rest on $\mathbf{I}$.

Now $\mathbf{P}$ can be realized on $U\subset M$ by a rotating platform, but
is obvious that at the same `time' on $U$, $\mathbf{I}$ cannot be realized
by any physical system.

$\mathbf{P}$ is a typical example of a reference frame for which it does not
exist a $(nacs|\mathbf{P})$ such that the time like coordinate of the frame
has the meaning of proper time registered by standard clocks at rest on $%
\mathbf{P}$.

\subsection{Sagnac Effect}

According to the classification of reference frames [20-22] it is indeed
trivial to see that $\mathbf{P}$ is \emph{not} proper time synchronizable or
even locally synchronizable.

This fact is\emph{\ not} very well known by the majority of physicists
(which are not specialists in \emph{RT})\emph{\ }and leads people from time
to time to claim that optical experiments done on a rotating platform \emph{%
disproves} \emph{SRT}. A recent claim of this kind has been done by Selleri
and collaborators on a series of papers [1-3] and Vigier [4] who thinks that
in order to explain the effects observed it is necessary to attribute a
non-zero mass to the photon and that this implies in the existence of a
fundamental reference frame.

There are also some papers that claim that in order to describe the Sagnac
effect it is necessary to build a non-abelian electrodynamics [5,6]. This is
only a small sample of papers containing wrong statements concerning the
Sagnac experiment. Now, the Sagnac effect is a well established fact (used
in the technology of the gyro-ring) that the transit time employed for a
light ray to go around a closed path enclosing a non-null area depends on
the sense of the curve followed by the light ray.

Selleri arguments that such a fact implies that the velocity of light as
determined by observers on the rotating platform cannot be constant, and
must depend on the direction, implying that nature realizes a
synchronization of the clocks which are at rest on the platform different
from the one given by the Einstein method.

Selleri arguments also that each small segment of the periphery of the
rotating platform of radius $R$ can be thought as at rest on an inertial
reference frame moving with speed $\omega R$ relative to the laboratory
(here modeled by $\mathbf{I}$).

In that way, if the synchronization is done \`{a} l'Einstein between two
clocks at neighboring points of the small segment, the resulting measured
value of the one way velocity of light must result constant in both
directions ($c=1$), thus contradicting the empirical fact demonstrated by
the Sagnac effect.

Now, we have already said that $\mathbf{P}$ is not proper time
synchronizable, nor is $\mathbf{P}$ locally synchronizable (in general), as
follows from the rigorous theory of time in Relativity [20-22].

However, for two neighboring clocks at rest on the periphery of a uniformly
rotating platform an Einstein's synchronization can be done. Let us see what
we get.

First, let $<\widehat{x}^{\mu}>$ be a $(nacs|\mathbf{P})$ given by
\begin{equation}
\widehat{t}=t,\quad\widehat{r}=r,\quad\widehat{\phi}=\phi-\omega t,\quad%
\widehat{z}=z   \label{18}
\end{equation}
In these coordinates $\mathbf{g}$ is written as
\begin{align}
\mathbf{g} & =(1-\omega^{2}\widehat{r}^{2})d\widehat{t}\otimes d\widehat {t}%
-2\omega\widehat{r}^{2}d\widehat{\phi}\otimes dt-d\widehat{r}\otimes d%
\widehat{r}-r^{2}d\widehat{\phi}\otimes d\widehat{\phi}-d\widehat{z}\otimes d%
\widehat{z}  \notag \\
& =g_{\mu\nu}dx^{\mu}\otimes dx^{\nu}   \label{19}
\end{align}

Now take the two clocks A and B, at rest on $\mathbf{P}$. Suppose they
follow the world lines $\rho$ and $\rho^{\prime}$ which are infinitely close.

According to the chronometric postulate (see, e.g., [21,22] the events on $%
\rho$ or $\rho^{\prime}$ can be orderly. We write, e.g., as usual $%
e_{1}<e<e_{2}$ to indicate that on $\rho$, $e$ is posterior to $e_{1}$ and $%
e_{2}$ is posterior to $e$.

Now let
\begin{equation*}
e_{1}=(x_{e_{1}}^{0},x_{e_{1}}^{1},x_{e_{1}}^{2},x_{e_{1}}^{3}),\quad
e_{2}=(x_{e_{2}}^{0},x_{e_{1}}^{1},x_{e_{1}}^{2},x_{e_{1}}^{3}),
\end{equation*}
\begin{equation}
e^{\prime}=(x_{e^{\prime}}^{0},x_{e_{1}}^{1},x_{e_{1}}^{2},x_{e_{1}}^{3}),%
\quad e=(x_{e}^{0},x_{e_{1}}+\Delta x^{1},x_{e_{2}}+\Delta
x^{2},x_{e_{3}}+\Delta x^{3}),   \label{20}
\end{equation}
$e_{1}$ is the event on $\rho$ when a light signal is sent from clock A to
clock B. $e^{\prime}$ is the event when the light signal arrives at clock B
on $\rho^{\prime}$ and is (instantaneously) reflected back to clock A where
it arrives at event $e_{2}$. $e$ is the event simultaneously on $\rho$ to
the event $e^{\prime}$ according to Einstein's convention and we have
[20,21,26]
\begin{equation}
x_{e}^{0}=x_{e^{\prime}}^{0}+\frac{g_{0i}}{g_{00}}\Delta x^{i}\neq
x_{e^{\prime}}^{0}   \label{21}
\end{equation}
We emphasize that Eq.(\ref{21}) does not mean that we achieved a process
permitting the synchronization of two standard clocks following the world
lines $\rho$ and $\rho^{\prime}$, $\emph{because}$ standard clocks in
general do not register the ``flow'' of the time-like coordinate $x^{0}$.
However, in some particular cases such as when $g_{\mu\nu}$ is independent
of time and for the specific case where the clocks are very near (see below)
and at rest on the periphery of a uniformly rotating platform this can be
done.

This is so because standard clocks at rest at the periphery of a uniformly
rotating platform ``tick tack'' at the same ratio relative to $\mathbf{I}$.
Once synchronized they will \emph{remain} synchronized. It follows that the
velocity of light measured by these two clocks will be independent of the
direction followed by the light signal and will result to be $c=1$ \ every
time that the measurement will be done. This statement can be \emph{trivially%
} verified and is in complete disagreement with a proposal of [9] that we
discuss in section 3.2. We now analyze with more details what will happen if
we try the \emph{impossible} task (since $\mathbf{P}$ is not proper time
synchronizable, as already said above) of synchronizing standard clocks at
rest on a rotating platform which is the material support of the reference
frame $\mathbf{P}$.

Suppose that we synchronize (two by two) a series of standard clocks (such
that any two are very close\footnote{%
Very close means that $l/R\ll1$, where $l$ is the distance between the cloks
and $R$ is the radius of the plataform, both distances being done in the
frame $\mathbf{P}$.}) at rest and living on a closed curve along the
periphery of a rotating platform. Let us number the clocks as $0,1,2,...,n$.
Clocks $0$ and $1$ are supposed ``be'' at the same point $p_{1}$ and are the
beginning of our experiment synchronized. After that we synchronize, clock 1
with 2, 2 with 3,... and finally $n$ with 0. From Eq.(\ref{21}) we get
immediately that at the end of the experiment clocks 0 and 1 will not be
synchronized and the coordinate time difference between them will be
\begin{equation}
\Delta\widehat{t}=-\oint\frac{g_{0i}}{g_{00}}d\widehat{x}^{i}=\oint \frac{%
\omega R^{2}}{1-\omega^{2}R^{2}}d\widehat{\phi}   \label{22}
\end{equation}
For $\omega R<<1$ we have $\Delta\widehat{t}=\pm2\omega S$ where $S$ is the
area of the rotating platform and the signals $\pm$ refer to the two
possible directions in each we can follow around the rotating platform.

The correct relativistic explanation of the Sagnac experiment is as follows.
Suppose (accepting the validity of the geometrical optics approximation [8])
that the world line of a light signal that follows the periphery of the
rotating platform of radius $R$ is the curve $\sigma:\mathcal{R}\supset
I\rightarrow\mathcal{M}$ such that $\sigma_{*}$ is a null vector. Using the
coordinate $\widehat{t}$ as a curve parameter we have
\begin{equation}
\mathbf{g}(\sigma_{*},\sigma_{*})=(1-\omega^{2}R^{2})(d\widehat{t}\circ
\sigma)^{2}-2\omega R^{2}(d\widehat{\phi}\circ\sigma)(d\widehat{t}\circ
\sigma)-R^{2}(d\widehat{\phi}\circ\sigma)^{2}=0   \label{23}
\end{equation}
Then
\begin{equation}
\frac{d\widehat{\phi}\circ\sigma}{d\widehat{t}}=-\omega\pm1/R   \label{24}
\end{equation}
Then, the coordinate times for a complete round are
\begin{equation}
\widehat{T}_{\pm}=2\pi R/1\mp\omega R   \label{25}
\end{equation}
where the signals $\pm$ refer to the two possible paths around the
periphery, with $-$ when the signal goes in the direction of rotation and +
in the other case. It is quite obvious that $\widehat{T}_{\pm}$ can be
measured by a single clock.

Now, observe that the length\footnote{%
This result follows at once from the defintion of the projection tensor
\textbf{h} (eq.(\ref{1})) which gives the metric of an instantaneous
observer. The length $L$ of the periphery given by eq.(\ref{26}) is just the
sum of the measurements done by different instantaneous observers along the
periphery of the disc. Note that the claim by Klauber [15] that the space
geometry of the disc is flat is in contradiction with the way measurements
are realized in \emph{RT}.} of the periphery as measured with a set of
standard rulers at rest on a rotating platform is
\begin{equation}
L=2\pi R(1-\omega^{2}R^{2})^{-1/2}.   \label{26}
\end{equation}
Being $\tau_{\pm}$ the proper times measured by standard clock at rest at
the periphery of the rotating platform, corresponding to $\widehat{T}_{\pm}$%
, we have
\begin{equation}
\tau_{\pm}=(1-\omega^{2}R^{2})^{1/2}\widehat{T}_{\pm}=L(1\pm\omega R).
\label{27}
\end{equation}
This equation explains trivially the Sagnac effect according to Special
Relativity.

Selleri [1] calls \ the quantities $c_{\pm}$
\begin{equation}
\frac{1}{c_{\pm}}=\frac{L}{\tau_{\pm}}   \label{28}
\end{equation}
the global velocities of light in the rotating platform for motions of light
in the directions of rotation ($-$) and in the contrary sense ($+$). \ He
then argues that these values must also be the local values of the \emph{%
one-way} velocities of light, i.e., the values that an observer would
necessarily measure for light going from a point $p_{1}$ in the periphery of
a rotating platform to a neighboring point $p_{2}$. He even believes to have
presented an ontological argument that implies that Special Relativity is
not true. Well, what is wrong with Selleri's argument is the following.
Although it is true that the global velocities $c_{\pm}$ can be measured
with a single clock, the measurement of the \ local one way velocity of
light transiting between $p_{1}$ and $p_{2}$ requires two standard clocks
synchronized at that points. Local Einstein synchronization is possible and
as described above gives a local velocity equal to $c=1$, and this fact
leads to no contradiction.

We observe at this point that the claims of [1-3] have also been discussed
properly by [13-14], and our analysis touch on some complementary aspects of
the problem. The analysis of [14] is specially rigorous and elegant being
based on an intrinsic approach to \emph{RT} which does not use reference
frames and that has been developed in detail in [27].

\subsection{Analysis of Claims by Chiu, Hsu and Sherry}

Another \textit{interesting} result, worth to be mentioned here because it
is a source of a misunderstanding is the following one that appears in a
paper by Chiu, Hsu and Sherry [9].

Take two clocks A and B (very close, i.e., as above $l/R\ll1$) and at rest
at the periphery of a rotating platform ($\mathbf{P}$ frame) with center at
the point with space coordinates ($0,R,0$) according to an $\mathbf{I}$
frame (the laboratory). Let
\begin{equation}
\mathbf{I^{\prime}}=\frac{1}{\sqrt{1-\omega^{2}R^{2}}}\partial/\partial t+%
\frac{\omega R}{\sqrt{1-\omega^{2}R^{2}}}\partial/\partial x   \label{29}
\end{equation}
be an inertial Lorentz frame moving with speed $\omega R$ relative to $%
\mathbf{I}$ and such that at the beginning of the synchronization procedure,
clock A is at rest relative to $\mathbf{I^{\prime}}$ and clock B is
differentially at rest relative to $\mathbf{I^{\prime}}$. It is assumed that
the platform is uniformly rotating in the xy-plane and clocks A and B are
synchronized when A is at the point with coordinates $(0,-\varepsilon,0,0$)
according to $\mathbf{I}$ and $(0,-(1-\omega^{2}R^{2})^{-\frac{1}{2}%
}\varepsilon,0,0)$ according to $\mathbf{I^{\prime}}$. Of course, clocks
synchronized in that way will not be synchronized according to the point of
view of an observer in the $\mathbf{I^{\prime\prime}}$ frame,
\begin{equation}
\mathbf{I^{\prime\prime}}=\frac{1}{\sqrt{1-\omega^{2}R^{2}}}%
\partial/\partial t-\frac{\omega R}{\sqrt{1-\omega^{2}R^{2}}}%
\partial/\partial x.   \label{30}
\end{equation}
This means that when the two clocks A and B are such that the coordinates of
A is $(0,\varepsilon,2R,0$) according to $\mathbf{I}$ , an observer at rest
in the $\mathbf{I}^{\prime\prime}$ frame comoving with the clock A \ and
such that at that event clock B is also at rest will see the clocks out of
phase if compared with their clocks that have been synchronized \`{a} l
'Einstein. This is so because $\mathbf{I^{\prime\prime}}$ is moving relative
to $\mathbf{I}$, and so clocks which are synchronized relative to $\mathbf{%
I^{\prime}}$ will not be synchronized according to $\mathbf{I^{\prime\prime}}
$. Now authors of [9] suppose that each segment (around the equator of
Earth) can be thought as at rest in an instantaneous inertial frame. Based
on this argument they claim that the velocity of light measured by the
stable clocks A and B synchronized as above described will suffer diurnal
variations. Of course, this is \emph{not} true. The observers at rest on
Earth will continue to find $c=1$ as the result of the measured one-way
velocity of light. And before leaving this section it is important to
emphasize that the supposition that a segment of the equator of Earth (or
the periphery of a rotating platform) is instantaneously at rest on an
inertial comoving frame, say $\ \mathbf{I}^{\prime}$, do not transform this
segment in a part of that frame. It continues to be part of the rotating
frame, say $\mathbf{P}$ and observers living in $\mathbf{P}$ can always know
that they are rotating, since rotation is an absolute concept in \emph{RT}.

We take the opportunity to call the reader's attention that the non
synchronizability of clocks on a rotating reference frame is crucial for an
understanding of the time as recorded by satellites (the GPS) and the time
registered by clocks at rest on the Earth, as described by Scorgie [28]. On
this issue see also [29].

Also, it is important to kept in mind that although we have speaked and
worked only with Einstein's synchronization in the above analysis, non
standard synchronizations can be also used. References on this important
issue are [30-35]

\subsection{Another \emph{nacs} to the Frame of a Rotating Platform.}

Some authors, e.g., [16-18] start the discussion of the rotating platform
problem by supposing that the \emph{natural }coordinate functions adapted to
a rotating platform must \emph{necessarily }be related with a ($nacs|\mathbf{%
I}$) by the following Lorentz like type transformations:
\begin{align}
\bar{t} & =t\cosh\Omega r-r\phi\sinh\Omega r,  \notag \\
\bar{r} & =r,\text{ }\bar{z}=z.  \notag \\
\bar{\phi} & =\phi\cosh\Omega r-\frac{t}{r}\sinh\Omega r,   \label{31}
\end{align}

These authors claim that these transformations are necessary and
the authors of [19] even claim that such coordinates are necessary
to solve some puzzles concerning the partial polarization of
electrons (or positrons) in a storage ring, for which they think
to have found an explanation by appealing to the Unruh-Davies
effects that according to quantum field theory must exist for some
kinds of accelerated motion. We defer the discussion of the last
problem to another paper. Here we are interested in a much simple
problem: which is the reference frame field $\mathbf{\bar{P}}$
such that the coordinates ($\bar{t},\bar{r},\bar{\phi},\bar{z}$)
are a (\textit{nacs|\textbf{P}})? The answer is:
\begin{equation}
\mathbf{\bar{P}=(}\cosh\Omega r)\partial/\partial t-(\frac{1}{r}\sinh\Omega
r)\partial/\partial\phi=\partial/\partial\bar{t}.   \label{32}
\end{equation}

Now, if we impose that $\mathbf{\bar{P}}$ is realized by a physical system
(as it is the case of the $\mathbf{P}$ frame) then the velocity of the
periphery of the platform with radius $r=\bar{R}$, must satisfy the
following constrain
\begin{equation}
\lim_{r\rightarrow\bar{R}}\sinh\Omega r=1.   \label{33}
\end{equation}

Now it is trivial to verify that $\alpha_{\mathbf{\bar{P}}}\mathbf{\wedge }%
d\alpha_{\mathbf{\bar{P}}}\neq0$ and then $\mathbf{\bar{P}}$ (like $\mathbf{P%
}$) is not proper time synchronizable. Moreover, the other characteristics
of the $\mathbf{\bar{P}}$ frame [given by the \emph{unique} decomposition
showed in eq.(1)] will be the same as for the $\mathbf{\bar{P}}$ frame if
\begin{equation}
\Omega=\frac{1}{r}\\arcsinh\omega\frac{r}{\sqrt{\left(
1-\omega^{2}r^{2}\right) }}   \label{34}
\end{equation}

In fact, with this identification $\mathbf{\bar{P}\equiv P}$ and this, of
course is the only satisfactory solution for our problem. Moreover, this
shows that to think that a Lorentz like type transformation [eq.(\ref{31})]
has some special significance is a nonsequitur and to think that these
coordinates help to solve the polarization problem of electrons in storage
rings is to forget that coordinates are labels, not the physics and to
mistaken the concept of coordinates charts with the concept of frames. In
\emph{RT} we can use any \emph{arbitrary} coordinate chart to do
calculations, but in that theory as well as in quantum field theory, in
general different reference frames are not \emph{physically equivalent}, a
concept that we study in [24].

\section{Conclusions}

In this paper we review some crucial issues associated with the concept of a
reference frame in $SRT$. By properly defining this concept and others
associated with it we investigated some issues concerning rotating reference
frames in Minkowski spacetime. We apply consistently the mathematical
notions of section 1, first to the study of the Sagnac effect in \emph{SRT}.
This has been done because it has been argued recently once again by
different people that Sagnac's effect cannot be explained by \emph{SRT}. We
disprove this statement by showing that this effect has a simple explanation
in that theory. Further we explore the problems which arise in
synchronization of standard clocks at rest on a rotating platform and the
problem of the measurement of the one way velocity of light in that frames,
clarifying many misconceptions that also appeared in recent literature. We
discuss also some claims that a more appropriate coordinate system [than the
canonical one given by eq.(\ref{18})] adapted to the frame that
mathematically models an uniforming rotating platform is given by a Lorentz
like transformation [eq.(\ref{31})] between the coordinates of the frame and
the ($nacs|\mathbf{I}$), where $\mathbf{I}$ is the inertial frame modelling
the laboratory. We show that such claims are equivocated.

\appendix

\section{Appendix A}

In this appendix we show the reason for the definition of a synchronizable
frame.

(i) We recall that a frame $\mathbf{Q}\in\sec TM$ is said to be
synchronizable iff there are $C^{\infty}$ functions $h,\bar{t}:M\rightarrow%
\mathfrak{R}$ such that $\alpha_{Q}=hd\bar{t}$ and $h>0.$

(ii) Now, we need to construct a proof for Proposition 1. Suppose that the
metric of the manifold in the chart ($U,\eta$) with coordinate functions $%
\langle x^{\mu}\rangle$ is $g=g_{\mu v}dx^{\mu}\otimes dx^{v}.$ Let $%
Z=(Z^{\mu}\partial/\partial x^{\mu})\in\sec TM$ be an arbitrary reference
frame and $\alpha_{Z}=g(Z,)=Z_{\mu}dx^{\mu},$ $Z_{\mu}=g_{\mu v}Z^{v}.$
Then, $g_{\mu v}(x)Z^{\mu}Z^{v}=1.$

Now, define
\begin{equation}
\begin{array}{ccccc}
\theta^{0} & = & (\alpha_{Z})_{\mu}dx^{\mu} & = & Z_{\mu}dx^{\mu}, \\
\gamma_{\mu v} & = & g_{\mu v}-Z_{\mu}Z_{v}; &  &
\end{array}
\tag{A1}  \label{A1}
\end{equation}

Then the metric $g$ can be written due to the hyperbolicity of the manifold
as
\begin{equation}
\begin{array}{rrr}
g & = & \theta^{0}\otimes\theta^{0}-\sum\limits_{i=1}^{3}\theta^{i}\otimes%
\theta^{i} \\
\sum\limits_{i=1}^{3}\theta^{i}\otimes\theta^{i} & = & \gamma_{\mu
v}(x)dx^{\mu}\otimes dx^{v}%
\end{array}
\tag{A2}  \label{A2}
\end{equation}

(iii) Now, suppose that there exist a coordinate chart ($V,\chi$), $U\cap
V\neq\emptyset$ with coordinate functions $\left\langle \bar{x}^{\mu
}\right\rangle $ such that $Z=\frac{1}{\sqrt{g_{00}}}\partial/\partial\bar
{%
x}^{0}.$ Then we can verify that
\begin{align}
\theta^{0} & =\sqrt{g_{00}(\bar{x})}d\bar{x}^{0}+\frac{g_{i0}(\bar{x})}{%
g_{00}(\bar{x})}d\bar{x}^{i},  \tag{A3} \\
\sum\limits_{i=1}^{3}\theta^{i}\otimes\theta^{i} & =\gamma_{\mu v}(\bar {x})d%
\bar{x}^{\mu}\otimes d\bar{x}^{v}  \notag \\
& =\left( \frac{g_{i0}(\bar{x})g_{j0}(\bar{x})}{g_{00}(\bar{x})}-g_{ij}(\bar{%
x})\right) d\bar{x}^{i}\otimes d\bar{x}^{j},\text{ }i,j=\text{1, 2, 3}.
\label{A3}
\end{align}

Now let us apply the results of (ii) and (iii) above a synchronizable frame
\textbf{Q} defined in (i). In the coordinate chart ($V,\chi$) with the
coordinate functions $\left\langle \bar{x}^{\mu}\right\rangle $ we take $%
\bar{x}^{0}=\bar{t}$. Since in these coordinates $\alpha_{Q}=hd\bar{t}$, it
follows that in these coordinates $\mathbf{Q}=\frac{1}{h}\partial/\partial
\bar{x}^{0}$ and $g_{00}(\bar{x})=h^{2}$.

From the first line of Eq.(A3) it follows that in this case $g_{i0}(\bar
{x}%
)=0$, i.e., the metric in the coordinates $\left\langle \bar{x}^{\mu
}\right\rangle $ is diagonal.

(iv) Now, recall Eq.(\ref{21}) from the text that gives the time like
coordinates of two events $e$ and $e^{\prime}$ that are simultaneous
according to Einstein's simultaneity definition. We see that in the
coordinates $\left\langle \bar{x}^{\mu}\right\rangle $ the events $e$ and $%
e^{\prime}$ are such that $\bar{x}_{e}^{0}=\bar{x}_{e^{\prime}}^{0}$. This
condition justify the definition of a synchronizable reference frame and
turns the definition a very intuitive indeed.

\section{Appendix B}

Here we explain the meaning of the definition of a locally synchronizable
reference frame $\mathbf{Q}\in\sec TM$, for which $\alpha_{Q}\wedge
d\alpha_{Q}=0$.

We recall the following:

(i) A 3-direction vector field $H_{x}$ in a 4-dimensional manifold $M$ for $%
x\in M$ is a 3-dimensional vector subspace $H_{x}$ of $T_{x}M$ which
satisfies a \emph{differentiability condition}. This condition is usually
expressed as:

(dfc1) For each point $x_{0}\in M$ there is a neighborhood $U\subset M$ of $%
x_{0}$ such that there exist three differential vector fields, $X_{i}\in\sec
TM$ ($i=$1, 2, 3) such that $X_{1}(x),X_{2}(x),X_{3}(x)$, is a basis of $%
H_{x},\forall x\in U,$ or:

(dfc2) For each point $x_{0}\in M$ there is a neighborhood $U\subset M$ of $%
x_{0}$ such that there exist a one form $\alpha\in\sec T^{*}M$ such that for
all $X\in H_{x}\Leftrightarrow\alpha(X)=0$.

(ii) We now recall Frobenius theorem [36]: Let be $x_{0}\in M$ and let be $U$
a neighborhood of $x_{0}$. In order that, for each $x_{0}\in M$ there exist
a 3-dimensional manifold $\Pi_{x_{0}}\ni x_{0}$ (called the integral
manifold through $x_{0}$) of the neighborhood $U$, \emph{tangent} to $H_{x}$
for all $x\in\Pi_{x_{0}}$ it is necessary and sufficient that [$X_{i},X_{j}$]%
$_{x}\in H_{x}$, (for all $i,j=1,2,3$ and $\forall x\in\Pi_{x_{0}}$) if we
consider condition (dfc1). If we consider the (dfc2) then a necessary and
sufficient condition for the existence of $\Pi_{x_{0}}$ is that
\begin{equation}
\alpha\wedge d\alpha=0.   \tag{B1}  \label{B1}
\end{equation}

(iii) Now, let us apply Frobenius theorem for the case of a Lorentzian
manifold. Let $\mathbf{Q}\in\sec TM$ be a reference frame for which $\alpha_{%
\text{\textbf{Q}}}\wedge d\alpha_{\text{\textbf{Q}}}=0$.

Then, from the condition for the existence of a integral manifold through $%
x_{0}\in M$ we can write,
\begin{equation}
\alpha_{\text{\textbf{Q}}}(X_{1})=\alpha_{\text{\textbf{Q}}}(X_{2})=\alpha_{%
\text{\textbf{Q}}}(X_{3})=0.   \tag{B2}  \label{B2}
\end{equation}

Now, since \textbf{Q} is a time like vector field, Eq.(B2) implies that the $%
X_{i}$ ($i=1,2,3$) are spacelike vector fields. It follows that the vector
field \textbf{Q} is orthogonal to the integral manifold $\Pi_{x_{0}}$ which
is in this case a spacelike surface.

Now the meaning of a synchronizable reference frame (which is given by
Eq.(B1) becomes clear. Observers in such a frame can locally separate any
neighborhood $U$ of $x_{0}$ (where they ``are'') in time$\times$space.

\section{Appendix C}

In this paper we prove the decomposition of a general reference frame $%
\mathbf{Q}=(Q^{\mu}\partial/\partial x^{\mu})\in\sec TM$ given by Eq.(\ref{0}%
) in the text.

To prove the decomposition, choose a chart ($U,\eta$) with coordinate
functions $\langle x^{\mu}\rangle$ and write
\begin{equation}
D\alpha_{\text{Q}}=Q_{\mu;v}dx^{\mu}\otimes dx^{v}.   \tag{C1}  \label{C1}
\end{equation}

Then, all we need is to verify that the functions $\omega_{\text{Q}_{\mu
v}},\sigma_{\text{Q}_{\mu v}}$ and $\Theta_{\text{Q}}$ given by Eq.(\ref{2})
of the text satisfy the required conditions and that
\begin{equation}
Q_{\rho;\tau}h_{\mu}^{\rho}h_{\nu}^{\tau}=Q_{\mu;v}-a_{Q_{\mu}}Q_{\nu},
\tag{C2}  \label{C2}
\end{equation}
where in Eq.(C2) $h_{\mu}^{\rho}$ are the mixt components of the projection
tensor \textbf{h} given by Eq.(\ref{1}) of the text.\medskip

\textbf{Acknowledgments:}One of the authors (MS) would like to thank TWAS
for the traveling grant and CNPq for the local hospitality at IMECC-UNICAMP,
SP Brazil. Authors are grateful to Professors U. Bartocci, Y. Bozhkov, J.
Vaz, Jr. and A. Saa and Drs. A. M. Moya and D. S. Thober for stimulating
discussions.\medskip

{\large References}\medskip

[1] F. Selleri, \emph{Found. Phys}. \textbf{26}, 641 (1996); \emph{FPL}
\textbf{10}, 73 (1997).

[2] J. Croca J. and F. Selleri, \emph{N. Cimento B} \textbf{114}, 447 (1999).

[3] F. Goy and F. Selleri, \emph{Found. Phys. Lett. }\textbf{10}, 17 (1997).

[4] J. P. Vigier,\emph{\ Phys. Lett. A} \textbf{234}, 75 (1997).

[5] T. W. Barrett , \emph{Ann. Fond. L. de Broglie }\textbf{14}, 37 (1989);
\textbf{15}, 143 (1989); \textbf{15}, 253 (1990); in A. Lakhatia (ed.),
\emph{Essays of the Formal Aspects of Electromagnetic Theory}, World
Scientific, Singapore Pub. Co., pp. 6- 86, 1993; in T.W. Barrett and D. M.
Grimes (eds.), \emph{Advanced Electromagnetism}, World Scientific, Singapore
Pub. Co., pp. 278-313, 1995.

[6] P. K. Anastasovski, et al (AIAS group), \emph{Found. Phys. Lett.}\textbf{%
12}, 579 (1999).

[7] M. G. Sagnac, \emph{Comp. Rend}. \textbf{507}, 708 (1913); \textbf{157},
1410 (1913).

[8] E. J. Post, \emph{Rev. Mod. Phys}. \textbf{39}, 475 (1967).

[9] C. B. Chiu, J. P. Hsu and T. N. Sherry, \emph{Phys. Rev. D} \textbf{16},
2420 (1977).

[10] A. Asthekart and A. Magnon, \emph{J. Math. Phys.} \textbf{16}, 341
(1975).

[11] J. Anadan, \emph{Phys. Rev. D} \textbf{24}, 338 (1981).]

[12] A. A. Logunov, and Yu. V. Chugreev, \emph{Sov. Phys. Usp.} \textbf{31},
861 (1988).

[13] G. Rizzi and A. Tartaglia, \emph{Found. Phys.} \textbf{28}, 1663
(1998); \textbf{12}, 179 (1999); \textbf{28}, 1663 (1998).

[14] T. Matolcsi, \emph{Found. Phys.} \textbf{28, }1685 (1998).

[15] R. D. Klauber, \emph{Found. Phys.} \textbf{11}, 405-483 (1998).

[16] G. Trocheries , \emph{Philos. Mag}. \textbf{40}, 1143 (1949).

[17] Takeno H., \emph{Prog. Theor. Phys}. \textbf{7}, 367 (1952).

[18] V. A. De Lorenzi and N. F. Svaiter, \emph{Found. Physics} \textbf{29},
1233 (1999); \emph{Int. J. Mod. Phys. A} \textbf{14}, 717 (1999); in Yu. N.
Gnedin, A. A. Grib, V. M. Mostepanenko and W. A. Rodrigues, Jr., (eds.),
\emph{Proc. of the Fourth A. Friedmann Int. Symposium on Gravitation and
Cosmology}, pp.295-309, IMECC-UNICAMP Publ., Campinas, 1999.

[19] W. A. Rodrigues, Jr., M. Scanavini and L. P. de Alcantara, \ \emph{%
Found. Phys. Lett.} \textbf{3}, 59 (1990).

[20] W. A. Rodrigues, Jr. and M. A. F. Rosa, \emph{Found. Phys}. \textbf{19,
}705 (1989).

[21] W. A. Rodrigues, Jr. and de E.C. Oliveira, \emph{Phys. Lett. A }\textbf{%
140}, 479 (1989).

[22] R. K. Sachs and H. Wu, \emph{General Relativity for Mathematicians},
Springer Verlag, N. York, Berlin, 1977.

[23] J. E. Maiorino and W. A. Rodrigues, Jr., What is Superluminal Wave
Motion?, \emph{Sci. and Tech. Mag}. \textbf{2}(2) (1999), see {\footnotesize %
http://www.ceptec.br/ \symbol{126}stm}.

[24] W. A. Rodrigues, Jr. and M. Sharif,\emph{\ Equivalence Principle and
the Principle of Local Lorentz Invariance}, \emph{Found. Phys}. \textbf{31}%
(12) (2001).

[25] Y. Choquet-Bruhat, C. Dewitt-Morette and M. Dillard-Bleick, \emph{%
Analysis, Manifolds and Physics} (revised version), North Holland Pub. Co.,
Amsterdam, 1982.

[26] L. D. Landau and E. M. Lifschitz, \emph{The Classical Theory of Fields }%
(4th revised english edition), Pergamon Press, Oxford, 1975.

[27] T. Matolsci, \emph{Spacetime Withouth Reference Frames}, Akad\'{e}miai
Kiad\'{o}, Budapest, 1993.

[28] G. C. Scorgie, \emph{Eur. J. Phys}. \textbf{12}, 64 (1991).

[29] T. B. Bahder, \emph{IEEE Trans. Aerospace and Electr. Syst.} \textbf{34}%
, 1350 (1998).

[30] H. Reinchbach, \emph{The Philosophy of Space and Time}, Dover, New
York, 1958.

[31] W. A. Rodrigues, Jr. and J. Tiomno, \emph{Found. Phys.} \textbf{15},
945 (1985).

[32] R. Mansouri, and R. U. Sexl, \emph{Gen. Rel. Grav.} \textbf{8}, 497
(1977).

[33] R. Anderson, I. Vetharaniam, G. E. Stedman, \emph{Phys. Rep.} \textbf{%
295}, 93 (1998).

[34] C. Leuber,K. Aufinger, and P. Krumm, \emph{Eur. J. Phys. }\textbf{13},
170 (1992)

[35] T. Ivezic, \emph{Found. Phys. Lett. }\textbf{12}, 507 (1999).

[36] T. Aubin, \emph{A course in Differential Geometry} (Graduate Studies in
Mathematics, Vol. 27) (American Mathematical Society, Providence, 2001).

\end{document}